\documentclass[aps,pra,reprint,nofootinbib,superscriptaddress,twocolumn,showpacs,showkeys,longbibliography,amsmath,amssymb]{revtex4-1}
\usepackage{graphicx}% Include figure files
\usepackage{dcolumn}% Align table columns on decimal point
\usepackage{bm}% bold math
\usepackage{braket}
\usepackage{subfigure}
\usepackage[colorlinks,bookmarks=false,citecolor=blue,linkcolor=red,urlcolor=blue]{hyperref}
\usepackage[english]{babel}
\usepackage{changes}
\usepackage{CJK}

\begin{document}
\title{Interaction between an impurity and nonlinear excitations in a polariton condensate}
\author{Chunyu Jia}
\affiliation{Department of Physics, Zhejiang Normal University, Jinhua, 321004, China}
\author{Zhaoxin Liang}
\email{The corresponding author: zhxliang@zjnu.edu.cn}
\affiliation{Department of Physics, Zhejiang Normal University, Jinhua, 321004, China}
\date{\today}

\begin{abstract}
Exploring the dynamics of a mobile impurity immersed in the field excitations is challenging, as it requires to account for the entanglement between the impurity and the surrounding excitations. To this end, the impurity's effective mass has to be considered as finite, rather than infinite. Here, we theoretically investigate the interaction between a finite-mass impurity and a dissipative soliton representing nonlinear excitations in the polariton Bose-Einstein condensate (BEC). Using the Lagrange variational method and the open-dissipative Gross-Pitaevskii equation, we analytically derive the interaction phase diagram between the impurity and a dissipative bright soliton in the polariton BEC. Depending on the impurity mass, we find the dissipative soliton colliding with the impurity can transmit through, get trapped, or be reflected. This work opens a new perspective in understanding the impurity dynamics immersed in the field excitations, as well as potential applications in information processing with polariton solitons.
\end{abstract}
\maketitle

\section{INTRODUCTION}

The motion of an impurity through a dynamical medium of field excitations is a fundamental problem. In his seminal paper~\cite{Landau1933,Landau1948}, Landau first studied an electron dressed by phonons. Since then, such impurity problem has appeared in different incarnations, such as the Kondo~\cite{Kondo} and Cherenkov~\cite{Bolotovskii2009,Carusotto2006,Seetharam2021} effects, the polaron physics~\cite{MahanBook}, and the Landau criterion~\cite{Astrakharchik2002,Wouters2010,Liang2008,Amelio2020,He2021} for the sound speed of a superfluid. At present, there are great interests and efforts in studying a mobile impurity in a quantum medium in diverse areas~\cite{Chevy2010,Massignan2014,Adlong2020,Schmidt2018,Jorgensen2016,Hu2016,Zoe2020}.

Central to understanding the dynamics of an impurity in a quantum many-body medium is to include the entanglement between the impurity and the surrounding excitations on a wide range of energy scales. To achieve this task, one needs to consider the impurity's effective mass as being finite, instead of infinite~\cite{Astrakharchik2002,Wouters2010,Liang2008,Amelio2020,He2021}. In addition, the excitations surrounding the impurity can be linear or nonlinear excitations. For instance, in ultracold quantum gases, the Bogoliubov modes are linear excitations, and dark (or bright) solitons are nonlinear excitations. Numerous theoretical studies~\cite{Chevy2010,Massignan2014,Adlong2020,Schmidt2018,Jorgensen2016,Hu2016,Zoe2020} have already been carried out to study the interaction mechanism between the impurity and the excitations. These studies, however, mainly involve linear excitations and the impurity with an infinite mass~\cite{Chevy2010,Massignan2014,Adlong2020,Schmidt2018,Jorgensen2016,Hu2016,Zoe2020} or finite effective mass~\cite{Vashisht2022}. Thus, it is highly desired to study the interaction mechanism between a quantum impurity with the finite effective mass and the nonlinear excitations, such as the soliton, which is not only a key ingredient in the effective field theory, but also plays an important role in the information processing~\cite{Flayac2013}. In this largely unexplored area, we will be interested in the interaction mechanism between an impurity and a moving bright soliton in the exciton-polariton Bose-Einstein condensate (BEC).

The exciton-polariton BEC has emerged as a novel platform for studying the impurity-related problems. In comparison with previous systems~\cite{Chevy2010,Massignan2014,Adlong2020,Schmidt2018,Jorgensen2016,Hu2016,Zoe2020}, which mainly concern equilibrium quantum medium, the polariton condensates have the fundamental novel aspects associated with their inherent non-equilibrium character and the strong nonlinearity. Firstly, because the polariton BEC is open-dissipative, the excitations of an homogeneous polariton condensate exhibit exotic properties. For instance, the linear excitations are provided by the diffusive Goldstone modes~\cite{Wouters2007,Littlewood2006,Byrnes2012,Xu2017}, with observable ghost branches of Bogoliubov excitations~\cite{Pieczarka2015}. This
have already triggered questions and studies on the definition of superfluidity and the characteristic observables in a nonequilibrium
context, e.g., extension of the standard Landau critical velocity has been proposed~\cite{Janot2013,Keeling2011,Van2014,Gladilin2016,Juggins2018,Amelio2020,He2021}. Novel kinds of nonlinear excitations have also been observed in recent experiments, such as the oblique dark solitons and vortices ~\cite{vortex1,vortex2,vortex3}, or bright spatial and temporal solitons~\cite{sky1}. Secondly, compared to the light-only solitons in the optical setups, the excitonic
component of the polariton leads to weaker diffraction and stronger inter-particle interactions, implying, respectively, tighter localization
and lower powers for nonlinear functionality. These appealing properties of polaritons can be used for quantum information processing~\cite{Flayac2013}, quantum computation and simulation~\cite{Sanvitto2016}. In particular, Ref.~\cite{Sich2012} has engineered dissipative bright polariton solitons, whose picosecond response time
makes them more useful
for ultrafast information processing than the
light-only solitons of semiconductor cavity
lasers. Thus, a timely question arises: In a non-equilibrium polariton BEC, what is the interaction mechanism between an impurity with the finite effective mass and a dynamical medium with nonlinear excitations?

In this work, we theoretically investigate the interaction between a finite-mass impurity and the dissipative bright soliton in a polariton BEC. By using the Lagrange variational method in the framework of the open-dissipative Gross-Pitaevskii equation, we analytically derive the interaction phase diagram. Depending on the impurity mass, we find that the dissipative soliton colliding with the impurity can have three fates, i.e., it can transmit through, get trapped, or be reflected. Our analytical analysis agrees well with the numerical simulations based on the open-dissipative Gross-Pitaevskii equation.

The rest of the paper is organized as follows. In Sec.~\ref{sec:2}, we present the model which describes a polariton condensate.
Furthermore, we derive the analytic expression of the interaction using the Lagrange variational method.
Sec.~\ref{sec:3} investigate the influence of the effective mass of the impurity on the interaction phase diagram between a soliton and an impurity in a polariton condensate, by means of direct simulation of  the motion equations of variational parameters and the Gross-Pitaevskii equation. Various interaction effects such as transmission, reflection, and trapping of
the soliton by a repulsive impurity are described and verified by direct simulations for equation.  Finally, Sec.~\ref{sec:4} provides a summary and conclusions for this research.

\section{The theoretical model and Lagrangian approach}\label{sec:2}

We consider an exciton-polariton BEC under nonresonant pumping, which is created in a wire-shaped microcavity~\cite{Wertz2010} that bounds the polaritons to a quasi-one-dimensional (1D) channel. In the mean field theory, the time evolution of the polariton field is governed by an effectively 1D  driven-dissipative GPE for the condensate order parameter $\psi(x,t)$, which is coupled to a rate equation for the density $n_{R}(x,t)$ of reservoir polaritons~\cite{Berloff2008,dark_polariton1,Xu2017,Yu2021,Sabari2022}, i.e.,
 \begin{eqnarray}
i\hbar\frac{\partial\psi}{\partial t}&=&\Big[-\frac{\hbar^{2}}{2m}\!\frac{\partial^2}{\partial x^2}+V_{\text{imp}}+g_{C}|\psi|^{2}+g_{R}n_{R}\nonumber\\
&+&\frac{i \hbar}{2}(Rn_{R}-\gamma_{C})\Big]\psi+P_{\text{ad}}(x)\psi,\label{psi}\\
\frac{\partial n_{R}}{\partial t}&=&P_{\text{incoh}}(x)-\left(\gamma_{R}+R|\psi|^{2}\right)n_{R}. \label{rate}
 \end{eqnarray}
 In Eqs.~(\ref{psi}) and (\ref{rate}), the $m$ is the effective mass of lower polaritons, $P$ is the off-resonant continuous-wave pumping rate, $\gamma_C$ and $\gamma_R$ denote the lifetimes of the condensate and reservoir polaritons, respectively, $R$ is the stimulated scattering rate of reservoir polaritons into the condensate,  $g_{C}$ characterizes the strengths of the polariton interaction, while $g_{R}$ denotes the interaction strength between the reservoir and the polaritons. The impurity potential~\cite{Berloff2008,Cristofolini2013} is $V_{\text{imp}}=-V_0\delta(x)$, with the strength $V_0$. The $P_{\text{ad}}(x)$ in Eq.~(\ref{psi}) and $P_{\text{incoh}}(x)$ in Eq.~(\ref{rate}) are the incoherent pumping rates on the condensate and reservoir~\cite{Sitnik2022}, respectively. The parameters $g_{C}$, $g_{R}$, and $R$ have been rescaled into the one-dimensional case by the width $d$ of the nanowire thickness, i.e., $g_{C}\rightarrow g_{C}/\sqrt{2\pi d}$, $g_{R}\rightarrow g_{R}/\sqrt{2\pi d}$, $R\rightarrow R/\sqrt{2\pi d}$. We aim to investigate the interaction mechanism between the impurity and the nonlinear excitations.

As the first step, let us determine the steady state of Eqs. (\ref{psi}) and (\ref{rate}), which will provide the density background for the nonlinear excitations. Following Ref.~\cite{dark_polariton1}, when the pumping rate $P$ in Eq. (\ref{rate}) exceeds the critical value $P_{\text{th}}=\gamma_R\gamma_C/R$, a stable condensate with the condensate density $n^0_C=(P_{\text{incoh}}-P_{\text{th}})/\gamma_C$ can be created. The corresponding steady-state reservoir density is $n^0_R=\gamma_C/R$, with $P_{\text{incoh}}=P_{\text{stat}}$.

By rescaling $\psi\rightarrow \psi/\sqrt{n^0_C}$ and denoting $m_R=n_R-n^0_R$, Eqs.~(\ref{psi}) and (\ref{rate}) can be recast into a dimensionless form as
 \begin{eqnarray}
i\frac{\partial\psi}{\partial t}&+&\frac{1}{2}\frac{\partial^2 \psi}{\partial x^2}+|\psi|^{2}\psi+\gamma\delta(x)\psi=2|\psi|^{2}\psi+\bar{P}_{\text{ad}}(x)\psi\nonumber\\
&+&(\bar{g}_{R}m_{R}+\frac{i}{2}\bar{R}m_{R})\psi,\label{Spsi}\\
\frac{\partial m_{R}}{\partial t}&=&\bar{P}_{\text{incoh}}(x)\!+\!\bar{\gamma}_{C}(1\!-\!|\psi|^{2})\!-\!\bar{\gamma}_R m_{R}\!-\!\bar{R}|\psi|^{2}m_{R}. \label{Srate}
\end{eqnarray}
Here $\bar{g}_{R}=g_{R}/g_{C}$, $\bar{\gamma}_{C}=\gamma_{C}\bar{\gamma}_{R}/\gamma_{R}$, $\bar{P}_{\text{ad}}(x)=P_{\text{ad}}/g_{C}n^0_C$ ,$\bar{P}_{\text{incoh}}=({P}_{\text{incoh}}(x)-P_{\text{stat}})/g_{C}n^0_C$ and $\bar{R}=\hbar R/g_{C}n^0_C$. The term with $\gamma=V_0/g_{C}n^0_C$ describes the impurity potential. Moreover, we have measured the time $t$ and the space coordinate $x$ in the units of $\tau=\hbar g n^0_C$ and $\xi=\sqrt{\hbar^2/m g n^0_C}$. Equations~(\ref{Spsi}) and (\ref{Srate}) are the starting point for our subsequent investigation of the interaction between the impurity and the nonlinear excitations in the polariton BEC. Note that the non-equilibrium nature of the model system is captured by the parameters of $\bar{R}$ in Eq. (\ref{Spsi}).

We are interested in the
fast reservoir limit, where the reservoir density in Eq. (\ref{Srate})
can be written as \cite{dark_polariton1}
\begin{equation}
m_R=\frac{\bar{P}_{\text{incoh}}(x)}{\bar{\gamma}_R}+\frac{\bar{\gamma}_{C}}{\bar{\gamma}_R}(1-|\psi|^{2}).\label{Frev}
\end{equation}
where $\bar{P}_{\text{incoh}}(x)=\bar{P}^c_{\text{incoh}}+\bar{P}^v_{\text{incoh}}(x)$, with the constant pumping rate $\bar{P}^c_{\text{incoh}}$ and the spatially dependent pumping rate $\bar{P}^v_{\text{incoh}}(x)$.
Following Ref.~\cite{dark_polariton1}, we insert Eq. (\ref{Frev}) into Eq. (\ref{Spsi}), and rewrite Eq. (\ref{Spsi}) as
\begin{eqnarray}
i\frac{\partial\psi}{\partial t}+\frac{1}{2}\frac{\partial^2 \psi}{\partial x^2}&+&\left|\psi\right|^{2}\psi\nonumber\\
&+&\gamma\delta(x)\psi= \frac{i}{2}[P(x)\!-\!\sigma-\chi\left|\psi\right|^{2}]\psi.\label{eq.1}
\end{eqnarray}
Here, we model $P(x)=\bar{R}\bar{P}^v_{\text{incoh}}(x)/\bar{\gamma}_R$ as a spatially modulated Gaussian function with the power $P_{0}$ and width $\omega$, i.e., $P(x)=P_{0}e^{-x^{2}/\omega^{2}}$;
the parameters $\sigma=-(\bar{P}^c_{\text{incoh}}+\bar{\gamma}_{C})\bar{R}/\bar{\gamma}_R$ and $\chi=\bar{R}\bar{\gamma}_{C}/\bar{\gamma}_R$ are referred to as the polariton loss rate and the gain saturation respectively. In deriving Eq. (\ref{eq.1}), the incoherent pumping of $\bar{P}_{\text{ad}}(x)$ is adjusted to be $\bar{P}_{\text{coh}}(x)=-2|\psi|^{2}-\bar{g}_{R}m_{R}$  within the current experimental capability~\cite{Cerna2009,Sitnik2022,WangY2021}. Below, we investigate the interaction between a bright soliton and the impurity as captured by the $\gamma$ term in Eq. (\ref{eq.1}).

Equation (\ref{eq.1}) can be viewed as a nonlinear Schr\"odinger equation subjected to a time-dependent perturbation of  the form $D(\psi)=i[P(x)-\sigma-\chi\left|\psi\right|^{2}]\psi/2$.
As a bench mark, let us recapitulate the unperturbed case $D(\psi)=0$ without the open-dissipative effects: (i) For vanishing nonlinearity in Eq. (\ref{eq.1}), Equation (\ref{eq.1}) can
be simplified into the linear Schr\"odinger equation with the delta-potential. It has the well-known exact solution $\psi_{\text{im}}(x)=\sqrt{\lambda}e^{-\lambda\left|x\right|}$ with $\lambda=\gamma$ that describes the impurity; (ii) For vanishing delta-potential $\gamma\rightarrow 0$, Equation (\ref{eq.1}) allows for the exact soliton solution $\psi_{\text{so}}=\text{sech}(\eta(x-c t))\exp(i(\eta^2-c^2)t/2+icx)$ with an arbitrary amplitude $\eta$.

Next, we take into account of the open-dissipative effects captured by $D(\psi)=i[P(x)-\sigma-\chi\left|\psi\right|^{2}]\psi/2$ in Eq. (\ref{eq.1}). Since the $\psi_{\text{im}}(x)$ and $\psi_{\text{so}}$ are no longer the exact solutions of Eq. (\ref{eq.1}), we exploit the Lagrangian approach of the perturbation theory to treat the open-dissipative effects. We assume a trial wave function as a combination of the bright soliton and impurity mode
\begin{eqnarray}
\psi\left(x,t\right)&=&\Big[\eta(t)\textrm{sech}\left[\eta(t) \left(x-z(t)\right)\right]e^{i\kappa(t) x}\nonumber\\
&+&a(t)\sqrt{\lambda(t)}e^{-\lambda(t)\left|x\right|+i\varphi(t)}\big]e^{i\phi(t)}, \label{eq.2}
\end{eqnarray}
where $\eta$, $z$, $\phi$, $\kappa$, $a$, $\lambda$ and $\varphi$ are the variational parameters. Specifically, $\phi(t)$ is the global phase of the trial wave function, $\eta (t)$ and $Z(t)$ are
the amplitudes and centre position of the bright soliton respectively, $\kappa(t)$ is referred to as the wavenumber of the soltion, $a(t)$ and $\lambda(t)$ are associated with the strength of the variable function induced by the impurity, $\varphi(t)$ is the relative phase between the soliton and impurity-induced function.

The key assumption underlying the ansatz~(\ref{eq.2}) is that the functional forms
of the soliton and the impurity-induced function are preserved in the presence of perturbation, whereas the corresponding parameters become slowly time-dependent. The time evolution of the parameters in Eq.~(\ref{eq.2})
can be obtained via the Euler-Lagrangian equations for the dissipative system~\cite{LagApp0,LagApp1,YilingZhang20501,ChunYuJia40502,Xu2019}
\begin{equation}
\frac{\partial L}{\partial q_{i}}-\frac{d}{dt}\left(\frac{\partial L}{\partial\dot{q_{i}}}\right)=2\mathcal{R}\left(\int_{-\infty}^{+\infty} D^{*}\left(\psi\right)\frac{\partial \psi}{\partial q_{i}}\right), \label{eq.3}
\end{equation}
with $\dot{q}_i\equiv dq_i/dt$ and $q_{i}={\eta, z, \phi, \kappa, a, \lambda, \varphi}$, and $\mathcal{R}$ labels the real part of the expression.
In Eq. (\ref{eq.3}), the Lagrangian $L=\int_{-\infty}^{+\infty}\mathcal{L} dx$ is referred as to the average Lagrangian of Eq. (\ref{eq.1}) with $D(\psi)=0$, where the Lagrangian density $\mathcal{L}$ is given by
\begin{equation}
\mathcal{L}=\frac{i}{2}\left(\psi^{*}\psi_{t}-\psi \psi_{t}^{*}\right)-\frac{1}{2}\left|\psi_{x}\right|^{2}+\frac{1}{2}\left|\psi\right|^{4}+\gamma\left|\psi\right|^{2}\delta(x). \label{eq.5}
\end{equation}

Inserting the ansatz (\ref{eq.2}) into Eq. (\ref{eq.5}), we calculate the average Lagrangian $L$ in Eq. (\ref{eq.3}) as
\begin{eqnarray}
L&=&-2\eta\dot{\phi}-2\dot{\kappa}z-a^{2}(\dot{\phi}+\dot{\varphi})+\frac{\eta^{3}}{3}-\kappa^{2}\eta-\frac{a^{2}\lambda^{2}}{2}+\gamma a^{2}\lambda\nonumber \\
&+&\gamma\eta^{2}\textrm{sech}(z)^{2}+2\gamma\eta a\sqrt{\lambda}\textrm{sech}(z)\cos(\varphi)+O\left(a^{4}\right).\label{Lagran}
\end{eqnarray}
Here we have ignored the higher-order terms of $O\left(a^{4}\right)$, as inspired by Ref. \cite{Forinash1994}. Physically, this corresponds to ignoring the direct interaction between the soliton and the local mode, except for the energy exchange through the defect. This approximation will be justified \textit{a posteriori} by comparing the analytical results from Eq.~(\ref{Lagran}) and the simulation results based on Eq.~(\ref{eq.1}).

By substituting Eq. (\ref{Lagran}) into Eq. (\ref{eq.3}), we obtain the equations of motion for the variational parameters $\phi$, $\kappa$, $\varphi$, $z$, $a$, $\eta$ and $\lambda$ in Eq. (\ref{eq.3}) as
\begin{widetext}
\begin{subequations}
\begin{eqnarray}
\dot{\eta}&=&\frac{1}{12}\Big[(6a^{2}+12\eta)(P_{0}-\sigma)
-12a\dot{a}-8\chi\eta^{3}-3\chi a^{4}\lambda\Big],\label{Dy1}\\
\dot{z}&=&\frac{1}{3}\left[3z\left(P_{0}-\sigma\right)-2\chi z\eta^{2}+3\eta\kappa\right], \label{Dy2}\\
\dot{a}&=&\frac{1}{4}\left[2a\left(P_{0}-\sigma\right)+4\gamma\textrm{sech}[z]\sin[\varphi]\eta\sqrt{\lambda}-\chi a^{3}\lambda\right], \label{Dy3}\\
%\dot{\kappa}&=&-\gamma\textrm{sech}[z]^{2}\tanh[z]\eta^{2}-\gamma a\cos[z]\textrm{sech}[z]\tanh[z]\eta\sqrt{\lambda}, \\
\dot{\kappa}&=&-\gamma\textrm{sech}[z]^{2}\tanh[z]\eta^{2}
-\gamma a\cos[z]\textrm{sech}[z]\tanh[z]\eta\sqrt{\lambda},\label{Dy4}\\
\dot{\varphi}&=&-\gamma\textrm{sech}[z]^{2}\eta-\frac{\eta^{2}}{2}+\frac{\kappa^{2}}{2}-\gamma a\cos[\varphi]\textrm{sech}[z]\sqrt{\lambda} 
+\gamma a^{-1}\cos[\varphi]\textrm{sech}[z]\eta\sqrt{\lambda}+\gamma\lambda-\frac{\lambda^{2}}{2}, \label{Dy5}\\
\dot{\phi}&=&\gamma\textrm{sech}[z]^{2}\eta+\frac{\eta^{2}-\kappa^{2}}{2}+\gamma a\cos[\varphi]\textrm{sech}[z]\sqrt{\lambda}, \label{Dy6}\\
0&=&a\left(\gamma-\lambda\right)+\gamma\eta\lambda^{-1/2}\cos[\varphi]\textrm{sech}[z].\label{Dy7}
\end{eqnarray}
\end{subequations}
\end{widetext}

Equations (\ref{Dy1})-(\ref{Dy7}) are the key results of this work, which describe the interaction of an impurity and a bright soliton in the polariton condensate.
Note that without the dissipation (i.e., $P_{0}=\sigma=\chi=0$), the above equations obviously reproduce the result of Ref.~\cite{Forinash1994}. According to Eqs.~(\ref{Dy1}),~(\ref{Dy2}) and (\ref{Dy3}), the non-equilibrium nature of the polariton condensates will directly affect the soliton's center position $z$ and its amplitude $\eta$, as well as the impurity's amplitude $a[t] \lambda[t]^{1/2}$. Since $\phi$ does not appear in Eqs.~(\ref{Dy1}) and (\ref{Dy2}), the relevant equations for our study, Eq. (\ref{Dy6}) for $\dot{\phi}$ is not important. Equation (\ref{Dy3}), on the other hand, is crucial because it shows that the moving soliton excites the local mode. Note, Eq. (\ref{Dy7}) without the soliton ($\eta=0$) gives the correct value $\lambda=\gamma$ for the spatial decay of the impurity mode.

\section{Interaction between an impurity and a bright soliton}\label{sec:3}

In the previous section, we have used the Lagrangian approach to analytically derive Eqs. (\ref{Dy1})-(\ref{Dy7}). Below 
we construct the interaction phase diagram by solving Eqs. (\ref{Dy1})-(\ref{Dy7}) and comparing the results with the exact numerical simulations of the
dynamics governed by Eq.~(\ref{eq.1}), supplemented with the initial function of Eq. (\ref{Dy2}).

Let us first specify the initial conditions of Eqs. (\ref{Dy1})-(\ref{Dy7}). We assume the soliton is initially at $z=-10$, far from the impurity at $z=0$. The initial amplitude and velocity of the soliton are chosen as $\eta=0.1$ and $\kappa=0.02$, respectively. For other parameters ($a$, $\lambda$, $\varphi$ and $\phi$), we set their initial values as $0$.

\begin{figure}[htb]
\centering
\includegraphics[width=1\columnwidth]{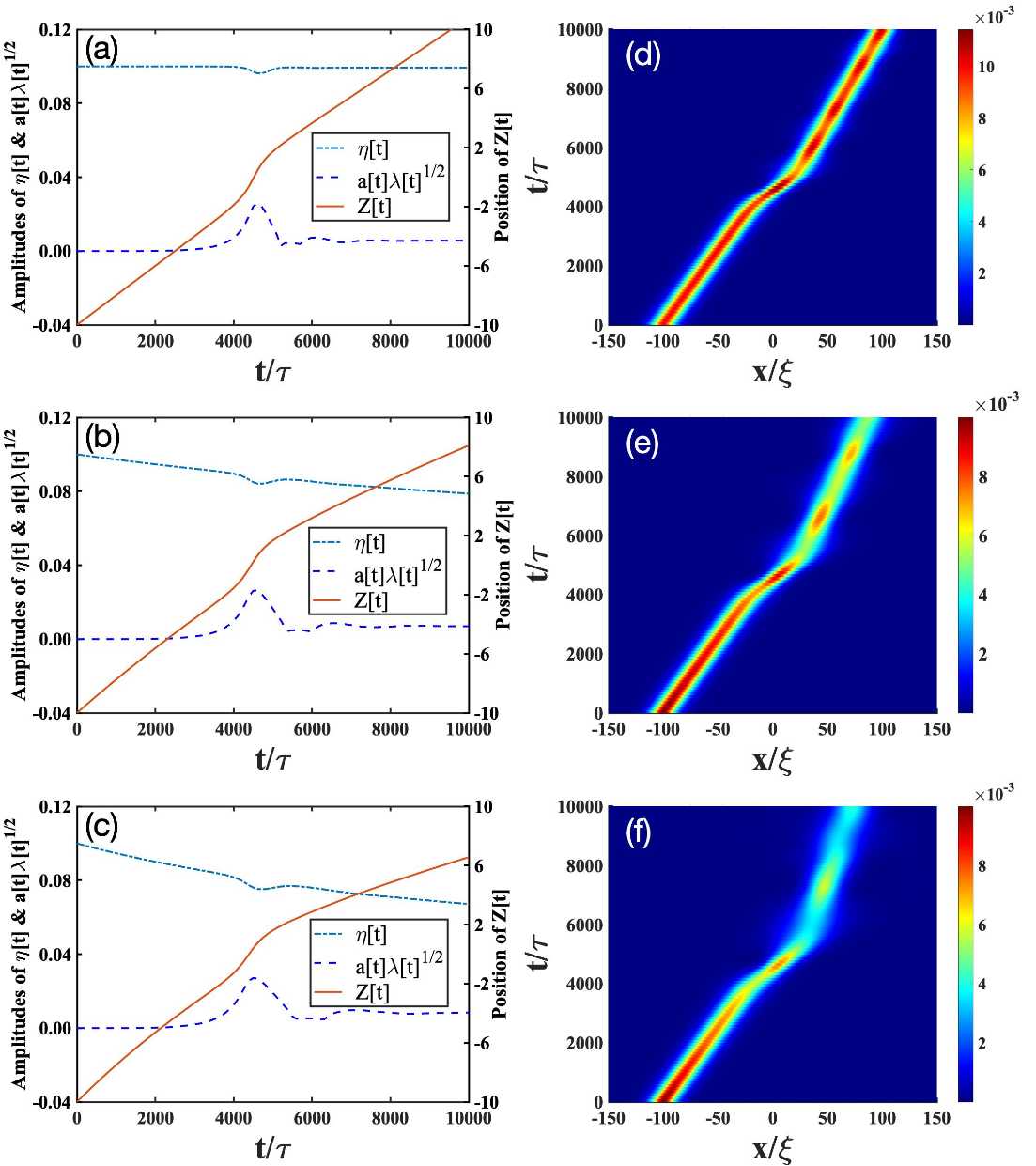}\\
\caption{Transmission scenario corresponding to the bright soliton with the initial value $\eta=0.01$ passing through the impurity with the strength of $\gamma=0.02$. The analytical results of Eqs. (\ref{Dy1})-(\ref{Dy7}) and the numerical simulation based on Eq.~(\ref{eq.1}) are plotted in the left and right columns respectively. In (a), (b) and (c), the position $Z$ of the bright soliton are plotted by  solid lines and scaled on the right axis; the amplitude $\eta$ of the bright soliton are plotted by dash-dotted lines and scaled on the left axis;  the impurity amplitude of $a\lambda^{1/2}$ are plotted by the dashed lines and scaled by the left axis. The other parameters are given as follows: $P_0=\sigma=\chi=0$ in (a) and (d); $P_0=\sigma=\chi=0.005$ in (b) and (e); $P_0=\sigma=\chi=0.01$ in (c) and (f).}
\label{figure1}
\end{figure}

\begin{figure}%[htb]
%\centering
\includegraphics[width=1\columnwidth]{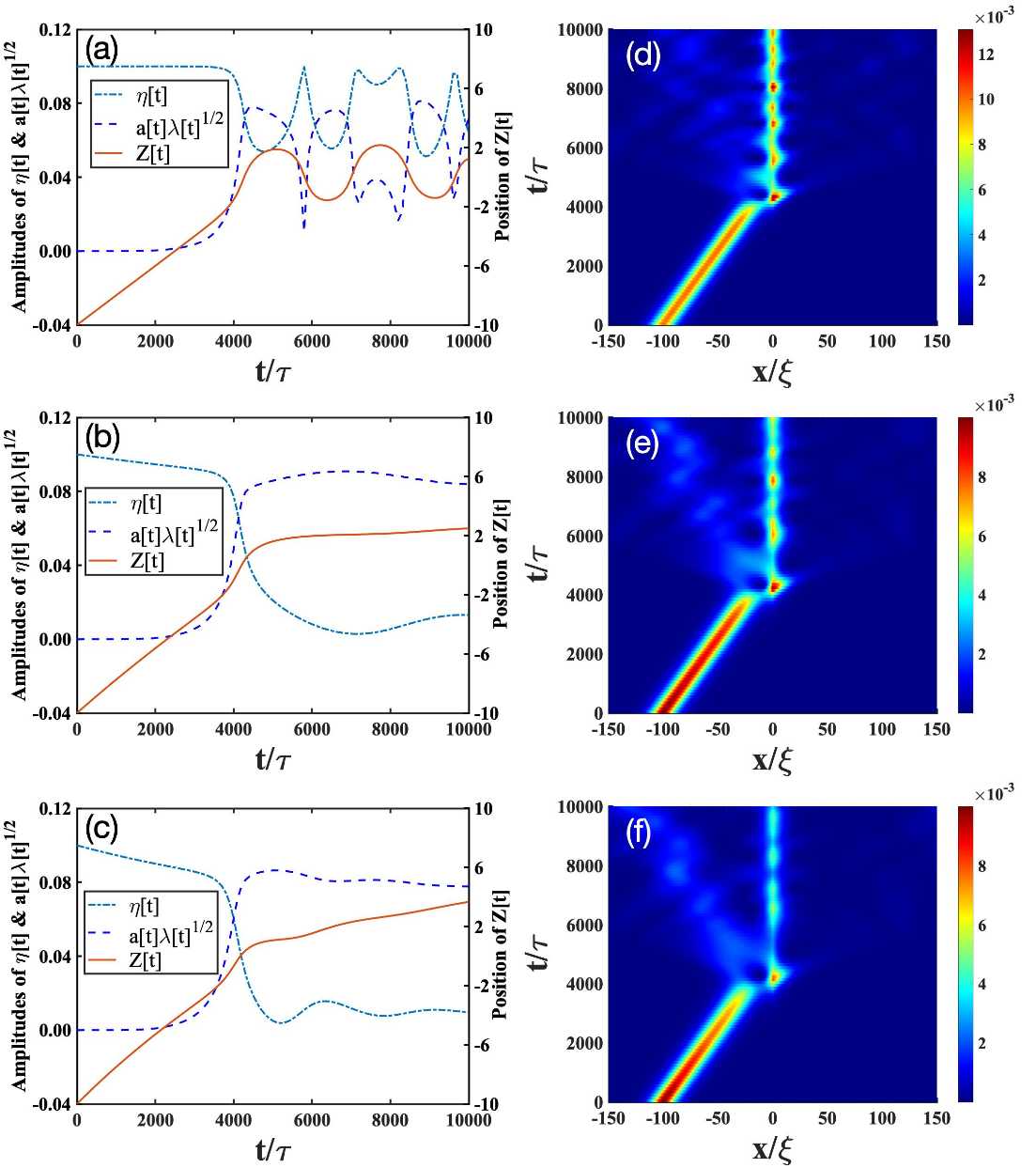}\\
\caption{Trapping scenario corresponding to the bright soliton with the initial value $\eta=0.01$ passing through the impurity with the strength of $\gamma=0.05$
The other parameters and descriptions about figures are same with the ones in Fig. \ref{figure1}.}
\label{figure2}
\end{figure}

\begin{figure}%[htb]
%\centering
\includegraphics[width=1\columnwidth]{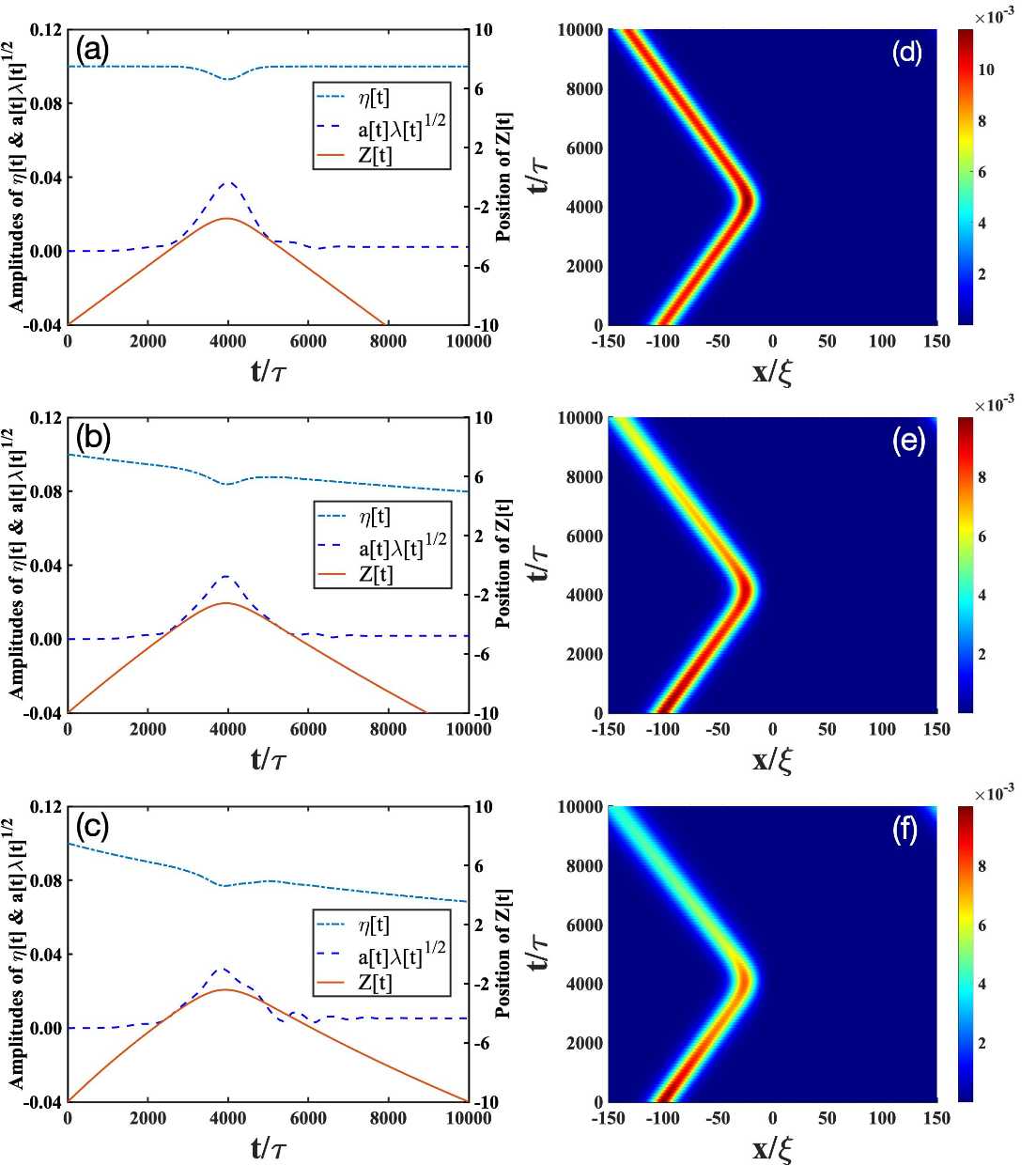}\\
\caption{Reflection scenario corresponding to the bright soliton with the initial value $\eta=0.01$ passing through the impurity with the strength of $\gamma=0.14$
The other parameters and descriptions about figures are same with the ones in Fig. \ref{figure1}.}
\label{figure3}
\end{figure}

We then solve the time-evolutions of the parameters $z$, $a$, $\kappa$, and $\varphi$ from Eqs.~(\ref{Dy2})-(\ref{Dy5}). The soliton amplitude $\eta$ is determined by Eqs. (\ref{Dy1})-(\ref{Dy3}), and $\lambda$ is calculated from Eq. (\ref{Dy7}). In addition, Eq. (\ref{Dy7}) allows us to
follow independently the evolution of the soliton and the impurity. The solutions to Eqs. (\ref{Dy1})-(\ref{Dy7}) are plotted in the left column of Figs.~\ref{figure1}, \ref{figure2}, and \ref{figure3}. To validate our variational approach, we also show the numerical results from the direct solutions of Eq. (\ref{eq.1}) on the right column of Figs.~\ref{figure1}, \ref{figure2}, and \ref{figure3}.

In understanding the interaction between the impurity and the quantum many-body medium, we emphasize the key role of the effective mass of the impurity~\cite{Carusotto2006,Seetharam2021}. For an infinite mass, corresponding to a pinned impurity~\cite{Astrakharchik2004,Carusotto2006}, a kinematic scale is set up by the sound speed of the superfluid according to the
Landau criterion. In contrast, an impurity with a finite mass is expected to recoil due to the interactions with the surrounding quantum gas, yielding novel physics beyond the
kinematic picture~\cite{Seetharam2021}. Indeed, quantum fluctuations become highly
relevant to the dynamics already for the slowly moving impurities with the finite mass.

Figures.~\ref{figure1}, \ref{figure2}, and \ref{figure3} show the interaction diagrams between the impurity and a bright soliton under various impurity trap strength $\gamma$, respectively. Different $\gamma$ corresponds to different effective mass $m_{\text{eff}}$ of the impurity~\cite{Forinash1994}. For $\gamma=0.02, 0.05, 0.14$ used in the plots, we have $m_{\text{eff}}=1.04, 1.10, 1.28$. In the following, we analyze how the impurity-soliton interaction is affected by the open-dissipative nature of the condensate, as captured by the parameters of $P_0$, $\sigma$ and $\chi$ in Eq. (\ref{eq.1}).

The results for $P_0=\sigma=\chi=0$ in the absence of dissipation~\cite{Forinash1994} are plotted in Figs.~\ref{figure1} (a), \ref{figure2} (a) and \ref{figure3} (a), respectively. Depending on $\gamma$, we find there exist three scenarios. (i) The transmission scenario [Fig.~\ref{figure1} (a)]: When $\gamma$ is small, the bright soliton directly transmits through the impurity. (ii) Trapping reflection scenario [Fig.~\ref{figure2} (a)]: When $\gamma$ increases, the bright soliton can be trapped by the impurity (see Fig. \ref{figure2} (a)). (iii) The reflection scenario [Fig.~\ref{figure3} (a)]: When $\gamma$ is strong enough, the bright soliton is reflected by the impurity.

To compare the interaction of the soliton with the impurity in the presence and absence of dissipations, we change the dissipative parameters of $P_0$, $\sigma$, and $\chi$ in each scenario:

(i) Transmission scenario. As mentioned before, in the absence of dissipation [Fig.~\ref{figure1} (a)], the bright soliton can simply pass through a light impurity ($m_{\text{eff}}=1.04$), almost unaffected by the latter. The dotted lines in Fig.~\ref{figure1} (a) denotes the amplitude of the impurity. There, the appearance of the maximal amplitude of the impurity indicates that the impurity mode can be excited during the collision with the bright soliton, but after the collision, the excitation returns to a very small level. This analysis is consistent with Fig.~\ref{figure1} (d) obtained from the numerical simulation of Eq.~(\ref{eq.1}). Thus we conclude that the analytical results in Eqs.(\ref{Dy1})-(\ref{Dy7}) not only provide a good solution to Eq. (\ref{eq.1}), but also allow us to follow independently the evolution of the bright soliton and the impurity. In the presence of dissipation, the amplitude of soliton gradually decreases after the collision with the impurity; see the solid lines in Figs.~\ref{figure1} (b) and (c). These results are consistent with the full numerical simulations in Figs.~\ref{figure1} (e) and (f). Comparing Figs.~\ref{figure1} (b) and (c), therefore, we see that the soliton amplitude decays faster when the dissipation parameter increases.

(ii) Trapping scenario. In the absence of dissipation [Figs. \ref{figure2} (a) and (d)], the bright soliton can be trapped by an impurity with a moderate mass ($m_{\text{eff}}=1.10$), as indicated by the position of the bright soliton [solid lines in Fig. \ref{figure2} (a)]. Furthermore, the impurity mode [dashed lines in Fig.~\ref{figure2} (a)] is strongly excited and begins to oscillate, whereas the soliton amplitude [dashed-dotted lines \ref{figure2} (a)] decreases drastically. This result is verified by the numerical simulations in Fig.~\ref{figure2} (d). In the presence of dissipation [Figs. \ref{figure2} (b), (c) and (e), (f)], the bright soliton can still be trapped by the impurity, but the oscillating behavior of the bright soliton begins to disappear. This can be understood, as the dissipation will destroy the low-energy excitations generated from the collisions of the bright soliton and the impurity.

(iii) Reflection scenario. In the absence of dissipation [Fig. \ref{figure3} (a) and (d)], the bright soliton can be reflected by a heavy impurity ($m_{\text{eff}}=1.28$). In contrast to the above transmission and trapping scenarios, dissipation has relatively small effects on the reflection scenario, as shown in Figs. \ref{figure3} (b), (e) and (d), (f). This can be expected, because the heavier the impurity is, the less excitations are created from the collisions.%Of course a sufficiently strong dissipation will inevitably lead to the destruction all the above three phenomena.

\section{CONCLUSION }\label{sec:4}

In summary, we have investigated the interaction dynamics of a soliton with an impurity mode
in the exciton-polariton condensates excited by a non-resonant pump.
Our study is based on the Lagrange variational approach, which allows us to analytically derive the equations of motion for each variational parameter. Depending on the interaction strength between the soliton and the impurity, we observe the occurrence of transmission, reflection, and trapping of the soliton by the impurity. We show that these effects are weakened with the increase of dissipation. Our analytical results of the interaction phase diagram agree well with the numerical results of the open-dissipative Gross-Pitaevskii equation. The present work goes beyond prior researches in the context of equilibrium systems, opening a new perspective toward understanding the non-equilibrium dynamics of a mobile impurity immersed in the field excitations.
\bigskip

We thank Yiling Zhang for stimulating discussions. This work was supported by the Zhejiang Provincial Natural Science Foundation (grant no. LZ21A040001), the National Natural Science Foundation of China (grant no. 12074344) and the key projects of the Natural Science Foundation of China (grant no. 11835011).

\bibliography{Reference}

\end{document}